# A Process for Reviewing Design Science Research Papers to Enhance Content Knowledge & Research Opportunities

Kweku-Muata Osei-Bryson
**VIRGINIA COMMONWEALTH UNIVERSITY**
Richmond, VA 23284, U.S.A.
Email: KMOsei@VCU.Edu

**Abstract:** Similar to most published Information Systems (IS) research, much of ICT4D research has involved the behavioral science research (BSR) category. One of the reasons for the dominance of the behavioral science research category is that this is the orientation of many IS doctoral programs (including those that train ICT4D researchers), and so often such programs do not involve technical IS courses. Without such technical content knowledge, many ICT4D researchers may not feel confident and competent to engage in *Design Science Research (DSR)* in such areas. An important question is how can such ICT4D researchers increase both their technical content knowledge and their intimacy with the DSR process? In this paper we describe a paper reviewing process for DSR that has the objectives: Increasing the reviewer's technical content knowledge of the given subject (e.g. Data Base Systems); increasing the reviewer's knowledge and understanding of approaches to designing and evaluating IT/IS artifacts; and enabling the reviewer to Identify opportunities for new DSR projects.

# 1. INTRODUCTION

Similar to most published Information Systems (IS) research, much of ICT4D research has involved the behavioral science research (*BSR*) category. Although design science research (*DSR*) papers have appeared in ICT4D journals (e.g. Namyenya, Daum, Rwamigisa, & Birner, 2022; Li, Thomas, Stoner, & Rana. 2020), there have also been calls for more *Design Science Research (DSR)* in the ICT4D space (e.g. Osei-Bryson & Bailey, 2019; Osei-Bryson, Brown & Meso, 2022). One of the reasons for the dominance of the *BSR* category is that this is the orientation of many IS doctoral programs, including those that train ICT4D researchers whether they are trained in countries with highly developed economies or those with developing economies. This also often means that the associated doctoral programs may not involve the student taking doctoral-level technical-oriented courses such database systems, decision support systems, knowledge management. Without such content knowledge, many ICT4D researchers may not feel confident and competent to engage in *DSR* in such areas. An important question is: *How can such ICT4D researchers increase both their technical content knowledge and their intimacy with the DSR process?*

It is possible that some readers may at this point have the question: "*How is design science relevant to ICT4D research?*" However, Bailey & Osei-Bryson (2018) in their editorial "Contextual reflections on innovations in an interconnected world: Theoretical lenses and practical considerations in ICT4D" noted that "*Key elements in the ongoing work towards sustainable development will be the ability of stakeholders in development initiatives to collaboratively and <u>effectively design, implement and evaluate innovations that will be adopted and utilized within specific development contexts and environments</u>*". Further, several ICT4D research projects have demonstrated how DSR methods could be beneficial in this regard (e.g. Barclay, 2014; van Biljon, Marais & Platz, 2017; Uwaoma & Mansingh, 2018; Rao & McNaughton, 2019; Namyenya, Daum, Rwamigisa, & Birner, 2022; Mushi, Serugendo & Burgi, 2023). Further discussion on this is provided later in this section.

In this paper we describe a *Paper Reviewing Process* for *DSR* papers that has been used in some of the content-oriented IS doctoral seminars (e.g. Database Systems, Decision Support Systems,

Knowledge Management) of an IS PhD program at a USA research university. These content-oriented IS doctoral all focus on the student gaining knowledge that contained in research papers rather than in textbooks. Depending on the objectives of the given seminar, different approaches are used to review research papers and different deliverables are expected of the given student.

Several published papers have aimed to provide guidance on how to do quality reviews (e.g. Lee, 1995; Hirschheim, 2008), but typically the focus of such guidance is not on the PhD student enhancing his/her content knowledge of the given subject (e.g. database systems) or better understanding how to do DSR. Rather their focus is on enabling the reviewer to provide quality feedback to authors and editors (e.g. Lee, 1995), or been oriented towards BSR (e.g. Hirschheim, 2008). This suggested the need for different type of a *paper reviewing process*.

The *Paper Reviewing Process* for DSR papers that is presented here involves the doctoral student doing a detailed analysis and dissection of the given DSR paper in order for him/her to obtain a deep understanding of the given research problem, its proposed solution (i.e. resulting artifact(s)), and the evaluation of the artifact(s). The major objectives of this exercise are:
1) Increase the student's content knowledge of the given subject (e.g. Data Base Systems, Knowledge Management, Decision Support Systems).
2) Increase the student's knowledge and understanding of the DSR process for designing and evaluating IT/IS artifacts.
3) Facilitate the student's Identification of opportunities for new DSR projects.

Figure 1 displays how we expect objectives 1 & 2 to be accomplished. It should also be noted that presented paper reviewing process does include steps that address objective 3. It should also be noted that although increasing the student's competence in doing quality reviews for academic journals is not an objective of our paper reviewing process, however, over the years several current and former students have reported that such increased competence has been an additional benefit.

**Figure 1: Paper Reviewing Process & Enhancement of Student's Personal Knowledge Base**

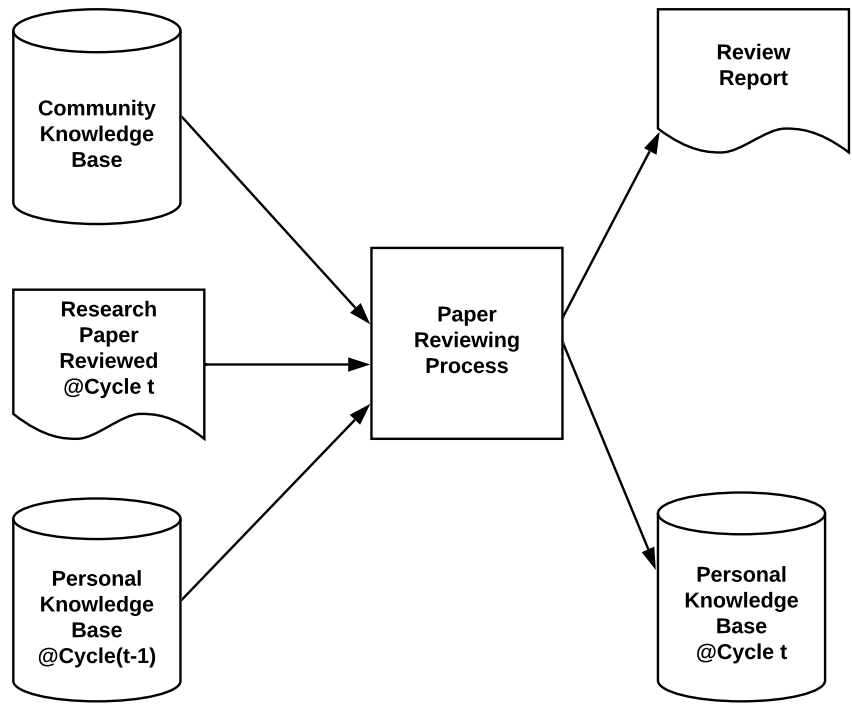

This paper reviewing process was initially used in 2002 in an IS doctoral seminar, and since then it has been enhanced, in part based on feedback from former and then current students, and also additional reflection. It has been used in some sections of the doctoral seminars in Database Systems, Knowledge Management, and Decision Support Systems since 2002, though other approaches are also used in the IS PhD program. The reader may note that while the paper reviewing process presented here has been used since 2002 that most of the referenced material is after 2002 and so the development of this approach is independent of material later presented in the referenced papers.

Gregor (2006) suggested that there are several theory types including Explanatory Theory, Predictive Theory, and Design & Action Theory ("*which says how to do something*"). Our proposed paper reviewing process belongs to the Design & Action theory category. It's aims to offer opportunities for ICT4D researchers to:

1) **Increasing their technical content knowledge of a given subject**.

The importance of ICT4D researchers being comfortable and competent with respect to the technical aspects of IS/IT (e.g. blockchain, data management, knowledge management, Big Data) in order to adequately address ICT4D societal problems is exemplified in several papers. For example,

- Blockchain Technology: Singh et al. (2021) focused on the use of blockchain technology "*to eliminate … inefficiencies and ensure trust*" in a public food distribution system that aimed at "*eradicating hunger and ensuring food security*";
- Data Management (Boakye, Zhao & Ahia, 2022): "*Despite the benefits of BCT in transforming Ghana's economy, there are still certain challenges and concerns to be resolved. … The second concern is data management, privacy, and security. Because information is maintained as a public ledger, confidentiality and security remain a concern for blockchains in general*";
- Big Data Qureshi (2021): "*Some would argue that the perverse effects of datafication for development are numerous ... Big Data is both a sociotechnical and epistemic process that involves the rapid combination, aggregation, and analysis of data in which researchers are tacitly accepting a commodification and quantification of knowledge*";
- Knowledge Management: Andoh-Baidoo, Ostuyi and Kunene (2014) focused on the need for a "*knowledge-management-system architecture on cybersecurity*" to address "*the cybercrime problem in SSA*".

However, as noted earlier, many IS doctoral programs do not include coursework in the technical areas of IS. This situation is further compounded by the fact that many of these programs do not require applicants to have master's level preparation in such technical areas. Given need for such knowledge in the ICT4D space, a relevant question is: *How can such ICT4D researchers without such academic preparation increase their technical content knowledge?* The *Paper Reviewing Process* that is presented in this paper is one of the means for addressing this question.

**2) Increasing their knowledge and understanding of the DSR process.**

Most ICT artifacts (e.g. including hardware, software, models, methodologies, frameworks) that are used in countries with developing economies were designed and developed in countries with advanced economies and typically are based on the values and perspectives of such societies. Yet as noted by Bailey and Osei-Bryson (2018) "*A sustainable world centers on supporting mechanisms involving the appropriate assessment of the context, suitable design and implementation approaches, the availability and use of appropriate resources, including relevant and suitable technologies, ICTs and digital innovations*". DSR can be used for the design, development, and evaluation of 'purposeful' artifacts that are contextually appropriate for ICT4D spaces. Several ICT4D researchers have confirmed the relevance and usefulness of DSR in ICT4D spaces including:

- Namyenya, Daum, Rwamigisa, & Birner (2022): "*applying the Design Science Research approach to this study provides insight into what it takes to design ICTs for development that address actual problems*";
- Mushi, Serugendo & Burgi (2023): "*the DSR method is used to design digital artifacts solutions that interact with the problem context and facilitate the development of an effective and efficient information system*";
- Van Biljon, Marais & Platz (2017): "*DSR offers an appropriate methodology to design information-based artifacts in a strategic and holistic way … the four-cycle DSR methodology provided structure and guidance in developing and evaluating an artifact to be used for development according to what can be described as a pragmatic stance*"

Although most published ICT4D research is in the *BSR* category, there have also been published DSR papers in a variety of domains, including cyber-security legislation development (Barclay, 2014), cyber-security architecture (Andoh-Baidoo, Osatuyi & Kunene, 2014), open knowledge repository (Van Biljon, Marais & Platz, 2017), pro-active healthcare (Uwaoma & Mansingh, 2018), disaster management (Rao & McNaughton, 2019), data justice and water governance (Hoefsloot et al., 2022); common challenges of small-farmers (Mushi, Serugendo & Burgi, 2023), agricultural extension services (Namyenya, Daum, Rwamigisa, & Birner, 2022).

Some ICT4D researchers have even customized DSR methodologies to be more appropriate for their research projects (e.g. Mushi, Serugendo & Burgi (2023): "*We customized the DSR method process by Peffers et al. (2007) from six to eight steps ...*"). To some extent this can be considered to be partial response to the suggestion of Osei-Bryson & Bailey (2019) that "*there should be research on developing ICT4D-oriented design science research methodologies*".

An increase in the proportion of published ICT4D research of the *DSR* category, will not only require welcoming journals, but also a pools of researchers and potential reviewers who are competent with respect to both ICT4D and *DSR*.   An important question is this: *How can ICT4D researchers increase their intimacy with the DSR process?* The presented *Paper Reviewing Process* is one of the means for addressing this question.

### 3)  *Facilitating the Identification of opportunities for new DSR projects.*

DSR focuses on both relevance and rigor. In ICT4D spaces, the relevance aspect relates to addressing important societal problems that involve one or more of the dimensions of development. Rigor can be considered to involve the rigorousness of the process for creating novel purposeful artifact(s) and associated theoretical and practical contributions to the body of knowledge, particularly with respect to ICT4D.

The presented Paper Reviewing Process involves the detailed analysis and dissection of the given DSR paper in order for him/her to obtain a deep understanding of the given research problem (including its elementary issues), its proposed solution artifact(s) including its solution components, and for each elementary issue to identify the solution component(s) of the artifact that addresses it & the effectiveness of the relevant solution components. The result is an enhancement of the users Personal Knowledge-base (see Figure 1) both respect to technical contain knowledge of the given IS subject area and also DSR process knowledge. This should increase the users competence in identifying and structuring important societal research problem, design appropriate solution(s), and select and appropriately apply relevant evaluation methods.

## 2. OVERVIEW ON DESIGN SCIENCE RESEARCH (DSR)

In this section we only intend to discuss the basic aspects of DSR that are relevant to reviewing DSR papers in order to accomplish the 3 objectives outlined in the Introductory section. March & Smith (1995) noted that: "*Whereas natural sciences and social sciences try to understand reality, **design science** attempts to create things that serve human purposes*". These created '*things*' are new knowledge plus at least one or more types of artifacts. Hevner et al. (2004) categorizes artifacts into the following four types: *Constructs*: vocabulary and symbols that provide the language in which problems and solutions are defined and communicated; *Models*: abstractions and representations; *Methods*: algorithms and practice; and *Instantiations*: implemented and prototype systems. Peffers et al. (2012) categorizes artifacts into six types: *Construct*, *Model*, *Framework*, *Method*, *Algorithm*, and *Instantiation* where: *Model*: Simplified representation of reality, documented using a formal notation or language; *Framework*: Meta-model; *Method:* Actionable instructions that are conceptual not algorithmic; and *Algorithm:* An approach, method, or process described largely by a set of formal logical instructions. Gregor and Jones (2007) offer a higher-level categorization of the types of artifacts: *Instantiations* or material artifacts (e.g. software, hardware, *Information System (IS)*, or "*the series of physical actions that lead to the existence of a piece of hardware, software, or an IS*"), *Theories* or abstract artifacts (e.g. *Construct*, *Model*, *Framework*, *Method*, *Algorithm*). They also note that humans use theories "*to guide the building of instantiations in the real world*" and that "*theory can be extracted from observation and inference from already instantiated artifacts*". In Table 1 we provide a list of artifact types that that have been described by these and other researchers including Vaishnavi & Kuechler (2004**).**

**Table 1: Types of DSR Artifacts**

| Artifact Type | Description |
|---|---|
| *Constructs* | The conceptual vocabulary of a domain |
| *Models* | Sets of propositions or statements expressing relationships between constructs |
| *Frameworks* | Real or conceptual guides to serve as support or guide |
| *Architectures* | High level structures of systems |
| *Methods* | Actionable instructions that are conceptual not algorithmic – How-To Knowledge |
| *Algorithm* | An approach, method, or process described largely by a set of formal logical instructions |
| *Instantiations* | Situated Implementations in certain environments that do or do not operationalize constructs, models, methods and other abstract artifacts |
| *Design Principles* | Core principles and concepts to guide design |
| *Design Theories* | A prescriptive set of statements on how to do something to achieve a certain objective. A theory usually includes other abstract artifacts such as constructs, models, frameworks, architectures, design principles and methods |

Several design science research methodologies (DSRM) have been presented by various researchers including Hevner et al. (2004), Peffers et al. (2007), Sein et al (2011), Drechsler & Hevner (2016), and Mullarkey & Hevner (2019). Some of these have high similarity to traditional system development methodologies to which IS students would have been previously exposed and so can be a convenient starting point for introducing students to the DSRMs of IS research. An example of such a DSRM is that of Peffers et al. (2007) which includes the following phases: *Identify Problem & Motivate, Define Objectives of a Solution, Design & Development, Demonstration, Evaluation*, and *Communication*.

## 3. DESCRIPTION OF THE PROPOSED PAPER REVIEWING PROCESS

### 3.1 A Model of the Design Science Research Problem (DSRP)

In this subsection we present our model (Figure 2) of a DSR problem. This model will be used in our proposed method for reviewing DSR papers. The *Issue* entity type could be considered to be the central element of this model as it is the *Issue*s that connect problem formulation (i.e. scope of the Formal Research Problem (FRP), the design of its proposed solution, and the adequacy of the evaluation of the proposed solution. This perspective is consistent with that of Peffers et al. (2007) who noted that: "*Because the problem definition will be used to develop an artifact that can effectively provide a solution, it may be useful to atomize the problem conceptually so that the solution can capture its complexity*". In the model represented by Figure 2 we view what we refer to as the formal research problem (*FRP*) as consisting of a set of *Issues*, some of which have Parent-Child links, and with the set of leaf-level *Issues* corresponding to the atoms of the *atomized conceptualization* of the research problem.

However, it should be noted that in some research papers the set of *Issue*s may not be clearly stated or even located in a single section of a given research paper. Again Peffers et al. (2007) noted: "*Some of the researchers **explicit**ly incorporate efforts to transform the problem into system **objectives**, also called meta-requirements or **requirements**, whereas for the others, these efforts are implicit as part of programming and data collection or **implicit** in the search for a relevant and important problem*". The sub-subsections below provide descriptions of the elements of this model.

**Figure 2: Model for Reviewing a Design Science Research Paper**

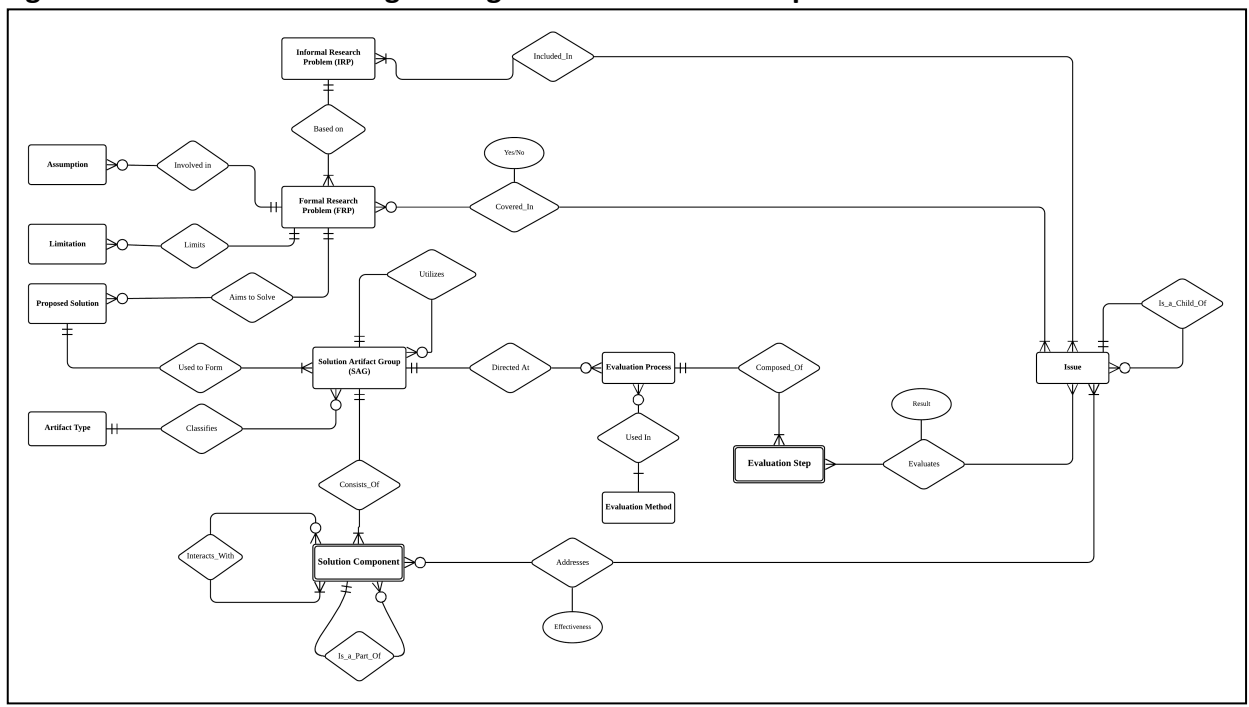

**3.1.1 The Issues:**

Within the context of DSR problem specification, terms such as *Requirement*, *Objective*, *Goal*, and *Criterion* have been used for what we refer to here as an *Issue*. These *Issues* (i.e. *Requirements, Objectives, Goals*, etc.) may be further decomposed by the authors into lower level sub-Issues (or sub-Requirements, Sub-Objectives, Sub-Goals, etc.) so there could be hierarchical relationships between the set of Issues (see Figure 3). Each *Issue* that is not further decomposed is considered to be a leaf-level *Issue*. An *Issue* at the highest level of the network of *Issues* can be considered to be a problem or opportunity that needs to be addressed.

**3.1.2 The Informal & Formal Research Problems**:

The given research problem can be described and defined in two ways: the problem that the authors may claim in their abstract and introduction to be addressing, and the problem that the authors are responsible for addressing based on their explicit and implicit assignment of *Issues* that fall within the scope of their research problem. The former is referred to here as *the Informal Research Problem* (*IRP*) while the latter is referred to as *the Formal Research Problem* (*FRP*). Table 2 provides a description of the *IRP*, the *FRP* and their constituent elements while Figure 3 provides a graphical description of the relation between the *IRP* and *FRP*. A major task of the reviewer is to determine what constitute the scope of the *FRP*. Evaluation of the Proposed Solution artifact(s) should be done with reference to the *FRP*, not the *IRP*.

**Table 2: The IRP, FRP**

| | |
|---|---|
| *Informal Research Problem (IRP)* | This is a relatively general description of the research problem that typically is broader than the research problem that is actually addressed in the given paper. |
| *Formal Research Problem (FRP)* | The actual research problem that the author(s) is(are) responsible for addressing. Its scope is defined by a set of *Issues* that are to be addressed, and the sets of author-specified *Assumption*s & *Limitation*s. The *FRP* covers a subset of the Issues that are associated with the *IRP*. |
| *Assumption* | o   May be explicitly described by the author(s) as an assumption.<br>o   Could be a previous result, or a previously proposed artifact or theory that is assumed by the author(s) to be valid & as such is utilized by the author(s) in the current paper. This may include what are referred to as '*kernel theories*' but it should be noted that not all DSR projects assume the necessity of '*kernel theories*'.<br>o   Note that it may be more appropriate to consider what might initially appear to be an *Issue* but which is a result from previous research to be an *Assumption* rather than as an *Issue*.<br>o   Something that is undoubtedly true could not be considered to be an Assumption |
| *Limitation* | o   Things Explicitly described by the author(s) as a Limitation or Constraint |

**Figure 3: Relationship between Informal Research Problem & Formal Research Problem**

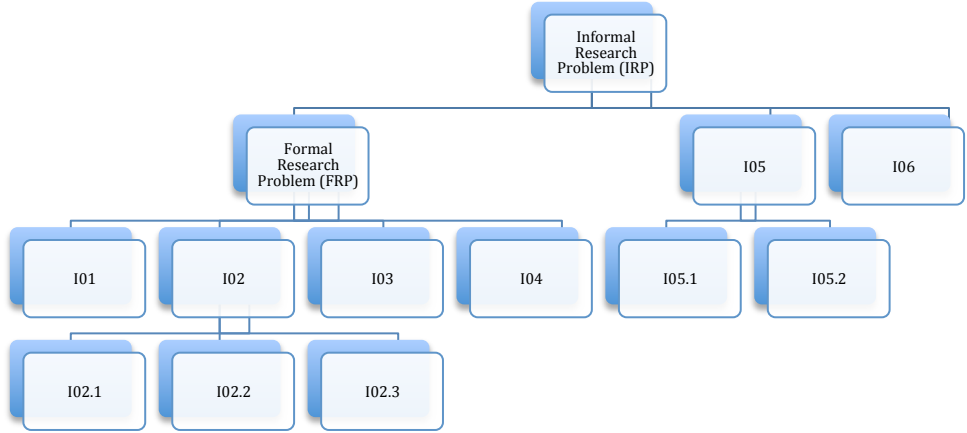

**Table 3a: Types of Issues within the Scope of the Formal Research Problem (FRP)**

| Type | Description |
|---|---|
| Direct (**D**) | o These are *Issues* directly articulated by the author(s) as being *included* within the *Scope* of the paper.<br>o In some cases the paper may include a single section in which the Issues/Requirements that fall within the Scope are formally identified as Issues/Requirements. In other cases **Issues**/Requirements may be located at various places in the paper, including the Introduction section.<br>o The authors may use the words: Requirements, Objectives, Goals, Criteria |
| Indirect Included (**I**) | o These are *Issues* indirectly articulated by the author(s) as being included within the *Scope* of the paper.<br>o Strengths & Limitations/Weaknesses of previous research by other researchers that the authors of the current paper discussed as being relevant to their research project. Each such Strength & Limitation/Weakness falls within the *Scope* of the Formal Research problem unless it has been explicitly excluded by the authors. |

**Table 3b: Other Issues that are included in the Informal Research Problem (IRP)**

| Type | Description |
|---|---|
| *Excluded* (**X**) | o Strengths & Limitations/Weaknesses of previous research by other researchers that the authors of the current paper discussed as being relevant to their research project but which they have **explicitly excluded** from the scope of their research project. |
| *Unidentified* | o Some of the *Direct* and *Indirect Issues* may have been decomposed by |

| Child Issues (**U**) | authors so that they have child *Issues*; in some cases the decomposition may be considered to be incomplete because some of the known child *Issues* of the given parent *Issue* were not identified by the authors. Such child *Issues* are considered to be *Unidentified Issues*. It should be noted that an *Unidentified* (U) Issue is a child Issue for which its parent is either a *Direct (D) Issue* or *Indirectly Included (I) Issue*. <br> For example it is known that the Security issue has as its children Confidentiality, Integrity and Availability. If Security and its children Confidentiality and Availability were included as *D* or *I* issues but no mention was made of the Integrity issue then Integrity would be considered to be an *Unidentified* (U) Issue. |
|---|---|
| Previously *Known* highly relevant but *Unmentioned* IS *Issues* (**K**) | o   Depending on the *IRP*, previously Known Information Systems (IS) *Issues* such as *System Performance* (e.g., processing time, storage cost), *User Acceptance* (e.g. Ease-of-Use, Usefulness, Satisfaction), *Group Issues* (e.g. consensus assessment, ability to produce group model), *Data Quality* (e.g. Accuracy, Consistency, Completeness, Credibility), *Data Usage Quality* may be highly relevant to the *IRP* and so could be considered to be included in the scope of the *IRP*, but which have not been mentioned or described in the paper. If any of these Issues is not a child of *D* or *I* Issue then it is considered to be a *K* Issue that is included in the scope of the *IRP* but excluded from the scope of the *FRP*. |

The reader may wonder what is the difference between a *Direct (D)*, *Indirect Included (I*) and *Excluded (X)* issues. For example, consider the statement from Chen and Lee (2003): "*According to Holsapple and Whinston, a multi-participant decision support system (MDSS) consists of four subsystems: (1) a language system which can handle both private and public messages; (2) a problem processing system (PPS) which is capable of knowledge acquisition, selection, and derivation; (3) a knowledge system which can store both private and public domain knowledge for multi- participants; and (4) a presentation system*". Each of the *Issue*s referred to in (1) – (4) would be considered to be an *Indirect Issue* if it was not mentioned by the authors as being objectives of the solution (i.e. *Direct*) Issues. However, given the statement "*The CDSS architecture shown .. is based on the MDSS framework by Holsapple and Whinston*" from Chen and Lee (2003) then these four (4) *Issues* would be considered to be *Direct (D) issues* unless any of them were explicitly excluded from the scope of the paper by the authors and thus considered to be a *X* Issue. So let us consider a case when these *Issues* were not mentioned by the authors as objectives of the solution. In this case they would all be considered to be *Indirectly Included (I)* Issues unless any of them were *explicitly excluded* from the scope of the paper by the authors and thus would considered to be an *X Issue*.

### 3.1.3 The Proposed Solution:

The overall solution artifact that has been proposed to address the *Issues* that fall within the scope of the *FRP* could be any of the possible artifact types (e.g. *Instantiation, Model, Method, Architecture*, etc.). Table 4 provides a description of the elements that are used to describe the proposed solution.

**Table 4: Proposed Solution and its Constituents**

| | |
|---|---|
| *Proposed Solution* | This is the overall solution that has been proposed to address the *Issues* that fall within the scope of the *FRP*. It consists of a set of one or more related *Solution Artifact Groups (SAGs)*. |
| *Solution Artifact Group (SAG)* | o  Each *SAG* may utilize elements of another *SAG*. For example, a *Model SAG* may utilize a set of *Construct* artifacts, or an *Instantiation* SAG is based on an *Architecture* or *Model SAG.*<br>o  Each *SAG* is not a component of any other *SAG*.<br>o  Each *SAG* has a specific artifact type, but its components could be of a single artifact type or of multiple artifact types. For example:<br>   ▪ A *Construct* type *SAG* would consist of a set of one or more *Constructs*,<br>   ▪ An *Architecture* type *SAG* (e.g. see Figure 6) could have components of different artifact types (e.g. *Method, Model*); while an *Instantiation* type *SAG* could be the software implementation of another SAG. |
| *Solution Component (SC)* | o  Each *SAG* consists of higher-level *SCs*, each of which in the paper, may have been recursively divided into lower level sub-components.<br>o  Each *SC* could be an artifact (e.g. *Model, Method, Construct*), or If the *SAG* is an *Algorithm* or a *Method* then each *SC* would be a step or sequence of highly related steps of the given algorithm or method. |

### 3.1.4 Evaluation Process:

Hevner et al. (2004) and Peffers et al. (2012) presented sets of evaluations methods that may be used for assessing the quality of IS artifacts (see Table 5a). However, it is also known that methods vary in their level of appropriateness for evaluating different artifact types. For example, while the *Expert Evaluation* method may be appropriate for evaluating a *Framework*, its appropriateness & persuasiveness in evaluating an *Algorithm* would not be high. Peffers et al. (2012) reported on the distribution of Evaluation Methods by Artifact types in several studies previous to 2012 (see Table 5b).

Multiple Evaluation Methods may be used for the direct or indirect evaluation of the solution artifact(s). For example, in our illustrative example the Logical Informed Argument and Expert Evaluation methods were used to evaluate the Architecture Artifact, while the Expert Evaluation method was used to evaluate the Instantiation artifacts.

**Table 5a: Some Evaluation Methods for DSR**

| Method | Description | Source |
|---|---|---|
| Logical Informed Argument | An argument with face validity. Use information from the knowledge base (e.g., relevant research) to build a convincing argument for the artifact's utility | *H, P* |
| Illustrative Scenario | Application of an artifact to a synthetic or real-world situation aimed at illustrating suitability or utility of the artifact | *H, P* |
| Analytical: Static | Examine structure of artifact for static qualities (e.g., complexity) | *H* |
| Technical Experiment | A performance evaluation of an algorithm implementation using real-world data, synthetic data, or no data, designed to evaluate the technical performance, rather than its performance in relation to the real world. | *P* |
| Controlled Experiment | Study artifact in controlled environment for qualities | *H* |
| Simulation Experiment | Execute artifact with artificial data | *H* |
| Testing: Black Box | Execute artifact interfaces to discover failures and identify defects | *H* |
| Testing: White Box | Perform coverage testing of some metric (e.g., execution paths) in the artifact implementation | *H* |
| Prototype | Implementation of an artifact aimed at demonstrating the utility or suitability of the artifact | |
| Subject-based Research | A test involving subjects to evaluate whether an assertion is true. | *P* |
| Expert Evaluation | Assessment of an artifact by one or more experts | *P* |
| Action Research | Use of an artifact in a real-world situation as part of a research intervention, evaluating its effect on the real-world situation | *P* |
| Observational: Case Study | Application of an artifact to a real-world situation, evaluating its effect on the real-world situation. | *H* |
| Observational: Field Study | Monitor use of artifact in multiple projects | *H* |
| Static Analysis | Examine structure of artifact for static qualities (e.g., complexity) | *H* |
| Architecture Analysis | Study fit of artifact into technical IS architecture | *H* |

| Optimization | Demonstrate inherent optimal properties of artifact or provide optimality bounds on artifact behavior | H |
| Dynamic Analysis | Study artifact in use for dynamic qualities (e.g., performance) | H |

*H*: Hevner et al. (2004);   *P*: Peffers et al. (2012)

**Table 5b: Evaluation Methods – Distribution in Some Previous Studies**

| All Journals | | | | | | | | | | IS Journals | | | | | | | | | |
|---|---|---|---|---|---|---|---|---|---|---|---|---|---|---|---|---|---|---|---|
| | Logical Argument | Expert Evaluation | Technical Experiment | Subject-Based Experiment | Prototype | Action Research | Case Study | Illustrative Scenario | none | Total | Logical Argument | Expert Evaluation | Technical Experiment | Subject-Based Experiment | Prototype | Action Research | Case Study | Illustrative Scenario | none | Total |
| Algorithm | 1 | | 60 | 1 | | | | 3 | | 65 | | | 4 | | | | | | | 4 |
| Construct | 3 | | 3 | 2 | 2 | | | 2 | | 12 | 1 | | | | 1 | | | | | 2 |
| Framework | 1 | 1 | | | 1 | | 1 | 4 | 1 | 9 | 1 | 1 | | | | | 1 | 1 | 1 | 5 |
| Instantiation | | | 5 | 1 | 1 | | | 1 | | 8 | | | 3 | | | | | 1 | | 4 |
| Method | 2 | | 14 | 4 | | | 7 | 6 | | 33 | 1 | | 2 | 2 | | | 6 | 1 | | 12 |
| Model | 3 | | 10 | | 2 | 2 | | 4 | | 21 | | | 1 | | 1 | 1 | | 1 | | 4 |
| Total | 10 | 1 | 92 | 8 | 6 | 2 | 8 | 20 | 1 | | 3 | 1 | 10 | 2 | 2 | 1 | 7 | 4 | 1 | |

Source: Peffers et al. (2012)

**3.2 Description of the Paper Reviewing Process for DSR Papers**

In this subsection a description of the paper reviewing process is provided (see Figure 4), and as well as a comparison of its steps to a DSRM (see Table 6). An illustrative example is provided in Section 4.

**Figure 4: Description of the Paper Reviewing Process**

| | Step | Description |
|---|---|---|
| 0 | Preparation | o Create a spreadsheet with 5 columns: *Label*, *Issue_Desc* (Description), *Issue_Text* (Extracted Supporting Text), *PageNumbers* (of Extracted Supporting Text), *Category* (i.e. *D, I, X, U, K*) |
| 1 | *Develop an Initial Overview Narrative of the IRP* | a) Do an initial reading of the paper.<br>b) Do a second reading of the *Abstract*, *Introduction* and *Conclusion* in order to get an initial understanding of the scope of the *IRP*.<br>c) Develop a brief narrative (e.g. a single paragraph) that provides an overview of your understanding of the *IRP*.<br>d) Condense the narrative above into a single sentence and place it in the *Issue_Desc* column of the first data row of the spreadsheet with *Label* "IRP" |

| 2 | Do Preliminary Identification of the Issues | a) Read the paper and extract text that appear to describe a specific Issue/Requirement, and place each such text into the *Issue_Text* column of its own row of the spreadsheet, and record the relevant page numbers in the corresponding *PageNumbers* column. |
| --- | --- | --- |
| | | b) Provide in a few words your initial description of each Issue & record this in the *Issue_Desc* column. |
| | | c) Do a preliminary categorization (i.e. *D, I, X*) of each such *Issue* & record in the *Category* column. |
| 3 | Review & Refine the Set of Issues | a) Review the preliminary set of Issues in order to identify if what is essentially the same Issue was described in multiple ways in the paper, or one Issue is the child of another. |
| | | b) Provide labels for the Issues beginning with 'I01' such that: if the same Issue was described in multiple ways then each such Issue would have the same label; if an Issue is a child Issue of another Issue (say I02) then its label would reflect this relationship (e.g. I02.01). |
| | | c) Sort the preliminary set of Issues based on the *Label* column. |
| | | d) Do any appropriate updating of the Category of each Issue, particularly the redundant ones. |
| | | e) Remove any redundant Issue. |
| | | f) Review each parent *Issue* to determine if its decomposition is complete or whether it has some child *Issues* (*U*) that were not identified by the author(s). |
| 4 | *Define the FRP* | a) Include the sets of *Direct Issues* (*D*), *Indirect Issues* (*I*), & *Unidentified Issues* (*U*) in the scope of the *FRP*. |
| | | b) Identify the set *Assumptions* & the set of *Limitations*, and include them in the scope of the *FRP*. |
| | | c) Include description of the author's definition for each important concept & term that is relevant to the definition of the *FRP*. |
| 5 | *Describe the IRP* | a) Review and analyze your narrative description of the IRP in order to determine any relevant previously known IS *Issue* (*K*) that are included in the scope of the *IRP*. Recall that a *K* Issue is one that is in the Knowledge Base but has not been explicitly discussed in the paper as a *D*, *I* or *X* Issue, and is not the unidentified child *Issue* (*U*) of a *D* or *I* Issue. |
| | | b) Include in the scope of the *IRP*, the Issues of the *FRP*, the set of *X Issues*, and the set of *K Issues*. |
| | | c) Develop a graphical description of the *IRP* that will also show its relationship to the Issues of the *FRP*. |
| 6 | *Provide your Assessment of the Quality of* | Assess the scope of the *FRP* within the context of the *IRP*. |
| | | o Is there any inconsistency within the set of *Issues* that are included in its scope? |
| | *the FRP* | o Are the *Assumption*s reasonable? Are any of the *Issues* included within the FRPs scope inconsistent with any of the *Assumptions*?<br>o Given its scope (i.e. *Issues included,* the *Assumptions/Limitations*), is the *FRP* a substantive research problem? Is it much different from other *FRPs* that have been adequately addressed? Is it similar to other previously identified significant *FRPs* that are currently 'begging' for an adequate *Solution*? |
| 7 | *Describe Authors' Justification of the FRP* | a) Describe the argument(s) if any that the author(s) use to justify the validity of the *FRP*.<br>b) Describe the argument(s) if any that the author(s) use to justify the current importance of the *FRP*. |
| 8 | *Provide your Description of the Proposed Solution* | a) Identify all proposed solution artifacts and *Solution Artifact Group*s that form the *Proposed Solution*.<br>b) For each *Solution Artifact Group* identify its corresponding *Solution Components*.<br>c) Describe in your own words each *Solution Artifact Group* and its corresponding *Solution Components*. For each **Solution Component (SC)**, identify the relevant the Page Number(s) & provide a substantive description that should be in your own words but may include relevant extracts from the text of the paper. Your use of extracts should not be excessive. |
| 9 | *Provide your Assessment of the Authors' Proposed Solution* | For each leaf-level *Issue*:<br>a) Identify the set of *SCs* that is (are) designed to address the given *Issue*.<br>b) Use Logical Arguments to provide your initial assessment of the level of effectiveness of the set of *SCs* in addressing the given *Issue*. |
| 10 | *Analyze the Authors' Evaluation of the Proposed Solution* | For each *Solution Artifact Group*:<br>a) Determine whether an appropriate evaluation method was used to evaluate the artifact.<br>b) Determine whether the selected evaluation method was applied appropriately<br>For each level-level *Issue*:<br>c) Identify whether the *Issue* was adequately evaluated.<br>d) Describe the result of the evaluation. |
| 11 | *Identify Research Opportunities* | Given your analysis in Step *6 (Assess the Quality of the FRP)*, Step *9* (*Provide your Assessment of the Proposed Solution*), and Step *10* (*Analyze the Authors' Evaluation of the Proposed Solution*):<br>a) Identify opportunities for future research based on limitations identified in these steps (i.e. 6, 9, 10);<br>b) Identify resources (e.g. knowledge, skills, hardware/software) that you would need to adequately address these opportunities in future |

| | | research. |
|---|---|---|
| 12 | *Describe Knowledge Gained* | Describe what you learnt during your review of this paper. |

**Table 6: DSR Paper Reviewing Process and DSRM**

| **Paper Reviewing Process Activity** | **Most related DSRM Activity** (Peffers et al., 2008) |
|---|---|
| Describe the Informal Research Problem (IRP) | Identify Problem & Motivate: Define Problem |
| Describe the Formal Research Problem (FRP) | Define Objectives of a Solution |
| Describe how Authors show the Importance of the FRP | Identify Problem & Motivate: Show Importance |
| Assessment of the Quality of the FRP | Not discussed |
| Describe Author(s) Justification of the FRP | Not discussed |
| Describe the Solution Artifact(s) | Design & Development |
| Provide your Initial Assessment of the Effectiveness of the Proposed Solution | Demonstration: Use Artifact to Solve Problem |
| Analyze the Authors' Evaluation of the Proposed Solution | Evaluation: Observe how effective, efficient |
| Identify Research Opportunities | Not discussed |
| Describe Knowledge Gained | Not discussed |

### 3.3 Complexity and Simplicity

Is there a simple process for doing a quality review of a paper, even if the only objective is to offer a high quality assessment of the quality of the paper as an input to journal editorial decision-making? Such a high quality assessment would require, at a minimum, that the reviewer reflects on the validity and relevance of each comment that he/she makes, and that the set of offered comments is internally consistent and also comprehensive with respect to the contents of the paper. This would require that the reviewer is sufficiently knowledgeable with respect to the given problem domain, and also has a process for ensuring validity, relevance, consistency, and comprehensive. It appears that in many cases reviewers do not use any such process.

A process can be described at the various levels of detail such as: only the *WHAT*s (the things are to be done), or only the *WHAT*s & *HOW*s (how each *WHAT* is to be done), or *WHAT*s, *HOW*s & *WHEN* (which involves the ordering of the set of *WHAT*s). A description of the level of only the *WHAT*s could at face value appear to be simple but would be ambiguous and in fact complex. A description at the level of the *WHAT*s, *HOW*s & *WHEN* may appear to be complex but would not be ambiguous. If the aim was to describe the process in a manner that provides guidance to a new user, it seems reasonable to expect that the ambiguity associated with a *WHAT*s only approach may not offer adequate guidance for a new user and attempts to apply it in practice would not be as simple as it might initially appear. A description at the level of the *WHAT*s, *HOW*s & *WHEN* would involve a greater level of detail than one involving only the *WHAT*s would in fact be simpler to apply.

The presented *Paper Reviewing Process* (*PRP*) involves the *WHAT*s, *HOW*s & *WHEN*. The reader may recall that this process has 3 solution objectives that go beyond providing quality input for a journal editorial decision-making. We submit that a quality reviewing process that focuses on our 3 objectives require disciplined attention, reflection and focus on validity, relevance, consistency and comprehensiveness.

There are processes that that are used in IS research & education that may initially appear to be complex processes to the student, and yet which over time the student learns. This includes research methodologies used in quantitative behavioral science research (*BSR*) that involves many activities, some of which are individually complex. Similarly research methodologies used in for qualitative *BSR*. For there are well known instructional approaches for processes that initially appear to be complex, particularly if the process is described as a sequence of steps. For example with respect to our presented *PRP*, this could sequentially focusing on different steps in different weeks, such as Steps 0 & 1, followed by Step 2, then Step 3, Steps 4 & 5, etc.

## 4. ILLUSTRATIVE EXAMPLE

For the illustrative example, we present an application of the 12-step *Paper Reviewing Process* to a paper (i.e. Chen & Lee (2003): *An Exploratory Cognitive DSS for Strategic Decision Making*) that has been used as a practice review assignment in one of the doctoral seminars. This paper was chosen for a several reasons including:

1) Although it is a DSR paper as it involves all phases of Peffers et al. (2007) DSRM (i.e. *Identify Problem & Motivate, Define Objectives of a Solution, Design & Development, Demonstration, Evaluation*, and *Communication*), it does not use the term '*design science*'. This is because DSR was being done by many IS and other researchers for decades before the formulation of DSRMs by IS researchers in the early 2000s. In fact it was the existence of such published research that motivated the development of such formal methodologies. It is important for entrants to DSR to be aware of this fact so that they can develop a fuller understanding of what is DSR, and not naively exclude papers that do not include the '*design science*' term.

2) The fact that though the topic of a paper does not explicitly refer to ICT4D, the research problem that it addresses may still be highly relevant to ICT4D contexts. For example, developing appropriate systems to support strategic decision is certainly highly important in ICT4D contexts.

3) The knowledge acquisition opportunities that a paper offers may still be highly relevant to the development of the knowledge base of ICT4D researchers although its topic does not explicitly refer to ICT4D. It is important that entrants to DSR not naively only focus on papers that appear to be about ICT4D.

**Steps 1 - 5: Identify set of Issues and Describe the IRP & FRP**

At this point it would be clear that I have differentiated the *FRP* from the *IRP*. The initial motivations for this was the result of 2 situations that often notice: 1) in many DSR papers that I was asked to review, the actual research problem that was addressed by the authors in their solution and/or evaluation was much more restricted than what was claimed by the authors; and 2) in their own research projects many doctoral students did not provide a rigorous definition of their own research problem and so had an unclear understanding of what their solution artifact should be doing, and what should be involved in an adequate evaluation. Further the description of the additional Issues that are in the *IRP* can lead to the identification of research opportunities.

While ideally there should be no overlap between any pair of leaf-level Issues, greater priority is given to the student providing a full coverage of all the Issues included in the scopes of the *FRP* and the *IRP*. The reader may recall that we consider the Issues that fall within the scope of the *FRP* include the *D*, *I*, and *U* Issues. An Issue that was previously identified in the community knowledge-base could only be an *I* Issue if the authors did not explicitly classify it as a *D* Issue, but had discussed it in their literature review and had not explicitly excluded it from the scope of their *FRP*. The aim here is for the student to be aware that when in the literature reviews of their research papers, he/she critiques prior research that he/she has brought to the attention of the reviewers Issues that the given reviewers may require to be addressed unless explicitly excluded from the scope of the student's *FRP*.

In Table 7a below be present a subset of the Issues that we identify as included in the scope of the *FRP* of our illustrative example; Appendix has table that includes all Issues that were identified as being in the scopes of the *FRP* and *IRP*. It should be noted that for the "Label" column of Table 7a hierarchical relationships are labeled using the format *"Ip", "Ip.q", "Ip.q.r"* which indicates that Issue "*Ip.q*" is a child Issue of "*Ip*", and Issue "*Ip.q.r*" is a child Issue of "*Ip.q*".

The table 7a reports an *Issue* (i.e. *I13.3: User Satisfaction*) as being an unidentified (*U*) *Issue*. How would this be determined by the student? Given that *Issues I13.1* (*Ease of Use*) and *I13.2* (*Usefulness*) are identified as being in the scope of the FRP, information from the User Acceptance knowledge-base (e.g. Maes and Poels, 2006) would suggest that these are child *Issues* of the User Acceptance *Issue*, and that the User Acceptance *Issue* includes other Issues such as *User Satisfaction* (i.e. I13.3)

**Table 7a: Example of Issues included in the Scope of the FRP**

| Label | Issue_Desc | Issue_Text | D/I/U/K | Page # |
|---|---|---|---|---|
| I01.1 | Surface Beliefs | providing tools to surface DM's tacit assumptions and beliefs | D | 148 |
| I01.2 | Detect Outmoded Model | provide direct aid in detecting and unlearning outmoded mental models | I | 148 |
| I01.3 | Unlearn Outmoded Models | provide direct aid in detecting and unlearning outmoded mental models | I | 148 |
| I01.4 | Enrich mental models | consciously helping enrich the … mental models | D | 149 |
| I01.5 | Validation & | facilitating mental model validation and | D | 149 |

|       | Integration          | integration                                                                                                      |   |     |
|-------|----------------------|------------------------------------------------------------------------------------------------------------------|---|-----|
| I02   | Support Creativity   | …. facilitating … creative thinking                                                                              | D | 148 |
| I03   | Support Forward Thinking | supporting the decision maker's … forward thinking                                                           | D | 149 |
| I04   | Support Backward Thinking | supporting the decision maker's backward … thinking                                                         | D | 149 |
| I05   | Support Problem Recognition | provide support in problem and opportunity recognition and diagnosis                                     | I | 149 |
| I11   | Security             | *handle both private and public messages; can store private domain knowledge*                                    | I | 152 |
| I11.1 | Confidentiality      | *handle … private … messages;*                                                                                   | I | 152 |
| I11.2 | Availability         | *capable of knowledge selection*                                                                                 | I | 152 |
| I11.3 | Integrity            | *refers to the prevention of unauthorized or improper data modification.*                                        | U | KB-Sc |
| I13.1 | Ease-of-Use          | provide ... information ... in a... user-friendly fashion                                                        | I | 147 |
| I13.2 | Perceived Usefulness | … provides a useful conceptual tool for decision makers                                                          | I | 148 |
| I13.3 | User Satisfaction    | "… *the extent to which users believe the information system available to them meets their information requirements …*'" | U | KB-UA |
| I13.4 | Responsiveness       | EIS provide ... information ... in a timely ... fashion                                                          | I | 147 |

KB:UA: Source is the community Knowledge-base on User Acceptance

Apart from the excluded (*X*) Issues, it is important for the student to identify those *K* Issues from the relevant community knowledge-bases that are highly relevant to *IRP* but which have not been identified as *D, I* or *U* Issues. The aim here is not for the student to do an exhaustive search of community knowledge-bases, but to do reinforcement learning by engaging Issues that he/she identified in previously reviewed papers of the current course as well as some Issues from related courses and subject areas that seem relevant to the *IRP*. For example if we consider that the description of the *IRP* is reflected in the statement: "*Research on decision support systems (DSS)/executive information systems (EIS) … has largely ignored the cognitive aspect of decision support. … this research emphasizes the need to support the decision maker's general thinking processes to reduce cognitive biases in decision making*" then Issues from the community knowledge base on DSS (e.g. Hosack et al., 2012), user acceptance (e.g. Maes and Poels, 2006), IS success (e.g. Delone & McLean, 2003) knowledge management (KM) and knowledge management systems (e.g. Kwan & Balasubramanian, 2003; Rao & Osei-Bryson, 2007; Yeoh & Koronios, 2010; Stewart & Osei-Bryson, 2013; Bunnell, Osei-Bryson & Yoon, 2019), security and others could be highly relevant. The student's exploration of such Issues from the student's current personal knowledge-base as well as the community knowledge-based as well as documenting what he/she has learnt from the detailed review of the current paper (see Step 12) offers the opportunity to strengthen and enhance his/her personal knowledge-base. This activity also offers the opportunity for the student to learn to rigorously define, understand and communicate the scope of his/her future *FRP*s. So in Table 7b, Issues 15.1-15.14 could have been obtained from Rao & Osei-Bryson (2007) and *I5.5* could have been obtained from Yeoh & Koronios (2010).

**Table 7b: Example of Issues excluded from the Scope of the FRP**

| Label | Issue_Desc | Justification for Inclusion of IRP | X/K | Page # |
|---|---|---|---|---|
| I14 | Evaluation & Choice | "*… instead of providing support for the evaluation and choice phase of the decision-making process*" | X | 149 |
| I15.1 | Currency | "*describes when the information was entered in the system*" | K | KB-KM |
| I15.2 | Credibility | "*describes the credibility of the source that provided the information*" | K | KB-KM |
| I15.3 | Transactional Availability | "*describes the percentage of time the information … is available due to the absence of update processes*" | K | KB-KM |
| I15.4 | Volatility | "*describes the time period for which the information is valid in the real world*" | K | KB-KM |
| I15.5 | Scalabilty & Flexibiliy | "*… easy expansion of the system to align it with evolving information needs*" | K | KB-KM |

KB-KM: Source is the community Knowledge-base (KB) on Knowledge Management

A FRP often involves some assumptions that are used in its framing. Such assumptions could include results from previous research that limit the scope of the Issues that the proposed solution has to address.

**Table 7c: Example of Assumptions included in the Scope of the FRP**

| Label | Assumption | Supporting Extracted Text | Page # |
|---|---|---|---|
| A01 | Focus should be on Problem/Opportunity identification | *A DSS for strategic decision making should focus on support in problem/opportunity identification instead of evaluation and choice phase of decision-making* | 149 |

| | | *process* | |
|---|---|---|---|
| A02 | | *"According to Holsapple and Whinston … a multi-participant decision support system (MDSS) consists of four subsystems …."* | 153 |

In our illustrative example while *Limitations* were discussed with respect to the instantiation artifact and also the evaluation of the artifact, there was no limitation that affected the scope of the FRP apart from the assumptions and the "X" Issue that was expressed by the authors.

**Table 7d: Example of Limitations included in the Scope of the FRP**

| Label | Limitation | Supporting Extracted Text | Page # |
|---|---|---|---|
| | ** None expressed *** | | |

The same term can have different meaning to different readers (including reviewers) and authors, and even by the same author in different papers. Similarly different terms could have the same meaning. Our aims here are: 1) for the student as they review the paper to be clear as to what the authors mean by their use of a given term; 2) for the student in the communication of his/her own research to do so in a manner that reduces the chance of readers misunderstanding

**Table 7e: Example of Description of the Concepts & Terms**

| Concept/Term | Author's Definition | Page # |
|---|---|---|
| Mental Model | *"… a mental model - a set of deeply held assumptions and beliefs - provides a useful conceptual tool for decision makers to simplify complex business environments and to impose order on volatile competitive conditions to reduce uncertainty"* | 148 |

**Figure 5: Graphic Description of Issues, IRP & FRP**

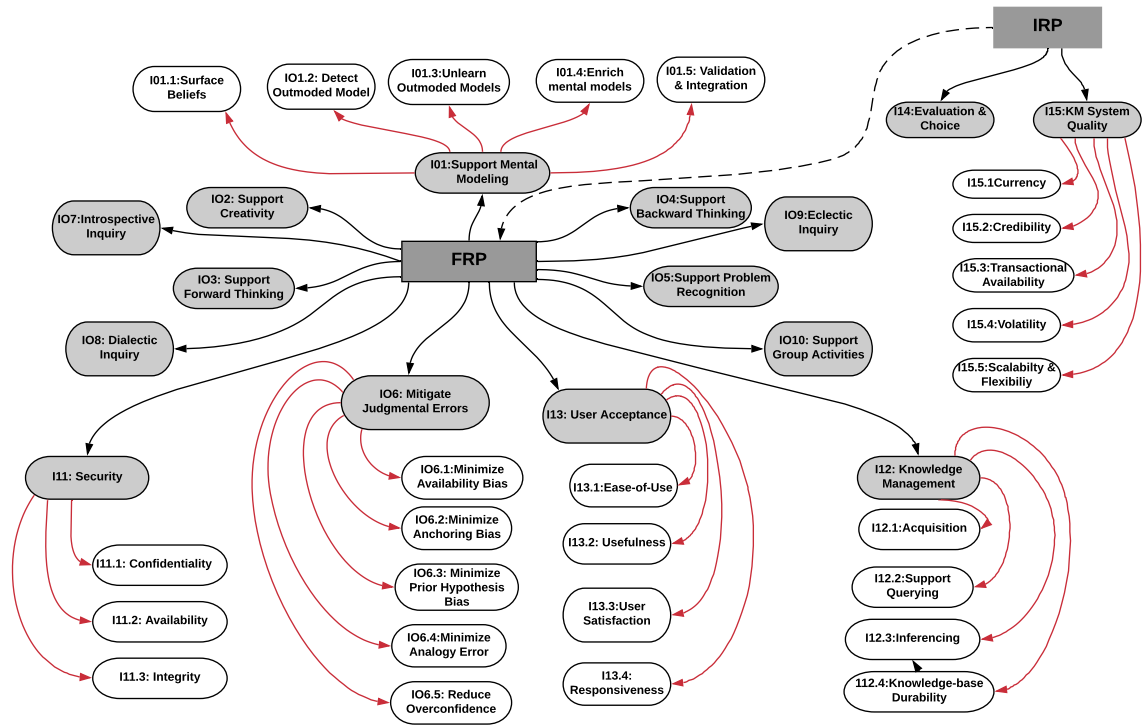

**Step 6: Provide your Assessment of the Quality of the FRP**

Given that the given *FRP* is a subset of the corresponding *IRP*, it is possible that even when the *IRP* could be considered to be a high quality research problem that the more restricted *FRP* may not be even if argued otherwise by the authors. Our aim in this step for the students to learn to assess the quality of potential *FRP*s, and so in their research they will select and define high quality FRPs. Table 8 provides an example of some questions and the student's possible responses, which could then to be used by the student to make an overall assessment of the quality of the *FRP*.

**Table 8: Example – Assessment of the Quality of the FRP**

| Some Questions to consider: | | |
|---|---|---|
| **Question** | **Y/N?** | **Your Reason** |
| a) Is there any Inconsistency between any pair of set of *Issues* that are included the scope of the *FRP*? | N | There is no inconsistency though there is the natural tension between the *Confidentiality* & *Availability* Issues. |
| b) Are the *Assumption*s reasonable? | Y | Based on recent research. |

| | | | |
|---|---|---|---|
| c) Are any of the *Issues* included within the scope of the *FRP* inconsistent with any of the *Assumptions*? | N | | |
| d) Given the scope of the *IRP*, is the *FRP* a substantive research problem? | Y | Although the scope of *FRP* does not include some important *Issues* of the *IRP* (e.g. Credibility, Volatility, Scalability & Flexibility), its scope includes most of the Issues of the *IRP*. | |
| e) Is the current *FRP* much different from other *FRPs* that have been adequately addressed? | Y | As noted by the authors: "*Various decision support systems provide their users quantitative modeling tools and easy data access. The cognitive aspect of decision support, however, has received relatively little research, although it has long been recognized as an important consideration of decision support systems design.*" | |
| f) Is the current *FRP* similar to other previously identified significant *FRPs* that are currently 'begging' for an adequate *Solution*? | N | At the time that the paper was written it addressed a significant but unaddressed *FRP* that are currently 'begging' for an adequate *Solution* (See above). | |

**Your Summary Assessment of the Quality of the FRP (based on your answers above):**

Based on the answers for the above questions, the *FRP* can be considered to be of good quality as it covers a substantive set of coherent Issues that are consistent with its set of reasonable Assumptions, and the *FRP* is a significant but unaddressed research problem.

**Step 7: Describe Authors' Justification of the FRP**

It always important for authors to establish the importance of their *IRP*; if the paper is being reviewed by rigorous reviewers who are aware of the difference between an *IRP* and the *FRP*, then it is important to establish the validity and importance of the *FRP*. In this step the student attempts to analyze the approach that has been used by the author(s) to establish the validity and importance of their *FRP*. The aim here is for the student to learn effective approaches for establishing the validity and importance of their own future *FRP*s.

**Table 9: Summary of Author's Justification of their FRP**

| What | Author's Justification (in your own words + relevant Verbiage Extract(s)) | Page # |
|---|---|---|
| *Validity* | Results from the community Knowledge-base was used to:<br>o Identify problems faced by strategic decision-makers (e.g. Executives may be blind-sighted by obsolete mental models which result in | 148 |
| | many business downfalls).<br>o Provide justification that addressing this is now feasible because of the "*increased understanding of managerial cognition*" and easier cognitive support provided by the then "new generation of computer technology".<br>o Scope the requirements for the DSS (e.g. *strategic decision making should focus on problem identification rather than evaluation and choice phase of decision-making process*) | |
| *Importance* | The authors noted that supporting the decision-makers cognitive processing is critically important but was previously not been adequately addressed in DSS research:<br>o "*Cognitive orientation or mental models play a very important role in a decision maker's understanding of business environments and ill-structured problems.*"<br>o "*The cognitive aspect of decision support … has received relatively little research, although it has long been recognized as an important consideration of decision support systems design*" | 147 |

**Step 8: Provide your Description of the Authors' Proposed Solution**

The presented proposed solution consists of three major artifacts, two of which are described graphically in Figures 6a and 6b, with the third being an instantiation of the architecture artifacts. The Conceptual Model (Figure 6a) was used to inform the design of the Architecture ((Figure 6b), and the Architecture was used to inform the construction of the Instantiation.

**Figure 6a: Model Artifact of the Illustrative Example**

Summary of the three-mode conceptual model

| Supporting mode | DSS supporting functions | Possible cognitive aid |
|---|---|---|
| Retrospective<br>● Recall past experience and cases.<br>   Analogical thinking. | Case Memory<br>● Aid storage, retrieval, and management of personal experience and cases. | ● Memory aid<br>● Reduce availability bias<br>● Analogical cases aid creative thinking. |
| Introspective<br>●Reflect on and examine the assumptions and belief system. | Cognitive Mapping<br>● Aid graphical representation of assumptions/belief systems (cognitive maps)<br>● Deduce the impact on specific construct through inference<br>● Manage and manipulate belief systems. | ● Surface and examine explicit and implicit assumptions.<br>● Overcome blind spots.<br>● Increase self-assurance. |
| Prospective<br>● Envision future state of business environments.<br>● Understand possible consequences of decisions. | Scenario Building<br>● Aid multiple scenarios construction process.<br>● Assist multiple scenarios management and manipulation. | ● Reduce overconfidence.<br>● Reduce anchoring effect.<br>● Reduce availability bias.<br>● Change frame of reference. |

Source: Chen, J. & Lee, S. (2003)

**Figure 6b: Architecture Artifact of the Illustrative Example**

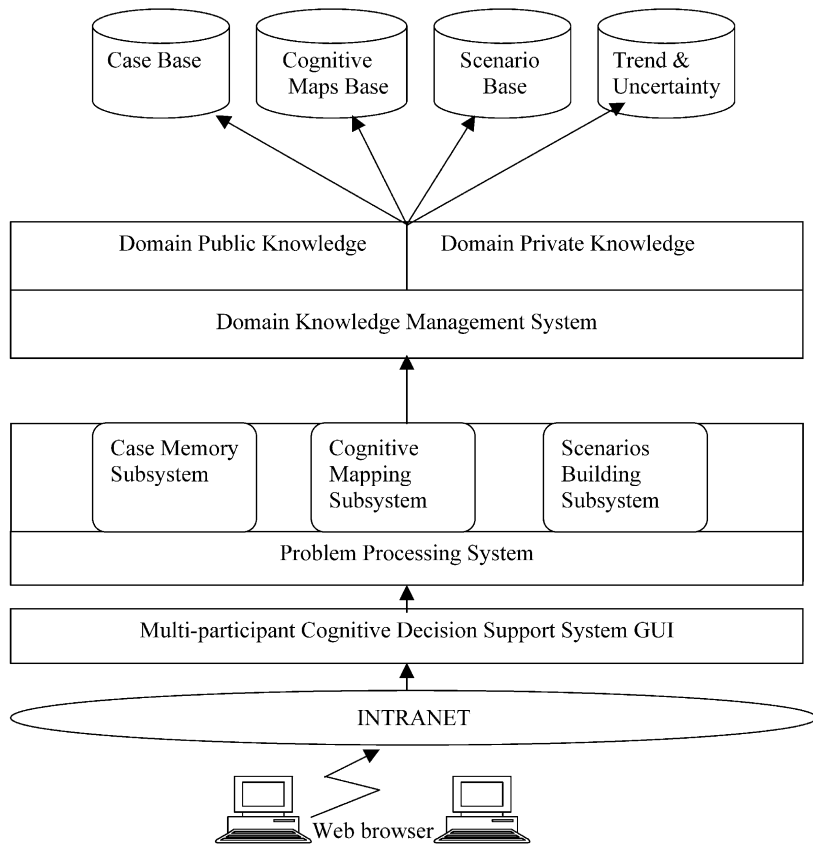

**Source**: Chen, J. & Lee, S. (2003)

The classification of these three major artifacts and the relationships between them would be described by the student (see Table 10a). The student would also be required to describe the proposed solution in his/her own words as well as the extracts that support his/her understanding. In Table 10b we present an example of the description of the solution components (*SC*s) of the Architecture artifact (i.e. SAG2) based on the graphical (e.g. Figure 6b) and narrative descriptions. It should be noted that in the "*SC Label*" column of Table 10b, hierarchical relationships are labeled using the format *"SAGx.SCp", "SAGx.SCp.q", "SAGx.SCp.q.r"* which indicates that Solution Component *"SAGx.SCp"* is a component of *"SAGx"*, *"SAGx.SCp.q"* is a sub-component of *"SAGx.SCp"*, and sub-sub-component *"SAGx.SCp.q.r"* is a sub-component of *"SAGx.SCp.q"*; Interaction relationships are presented in "*Interacts*" with column where the relevant SCs are listed.

**Table 10a: Solution Artifact Groups (SAGs)**

| SAG Label | Artifact Type | Description in your own Words | Page # | Utilized SAGs |
|---|---|---|---|---|
| SAG1 | Model | 3-Mode Conceptual Model | 151 | |
| SAG2 | Architecture | Architecture of Group Cognitive DSS | 152-156 | SAG1 |
| SAG3 | Instantiation | Instantiation of Group Cognitive DSS | 154-155 | SAG2 |

**Table 10b: Example of Solution Components (SCs)**

| Solution Artifact Group (SAG): SAG2 | | Artifact Type: Architecture | |
|---|---|---|---|
| SC Label | Description in your own Words | Interacts with | Page # |
| *SAG2.SC1* | User Interface / Presentation System (subsection 4.2) | *SAG2.SC2* | 152 |
| | | | |
| *SAG2.SC2* | Problem Processing System | *SAG2.SC1, SAG2.SC3* | 152-156 |
| *SAG2.SC2.01* | Case Memory Subsystem | *SAG2.SC3.01* | 152-153 |
| *SAG2.SC2.01.1* | Classify New Case | | 152 |
| *SAG2.SC2.01.2* | Record New Case | | 152 |
| *SAG2.SC2.01.3* | Retrieve Case | | 152 |
| *SAG2.SC2.02* | Cognitive Mapping Subsystem | *SAG2.SC3.02* | 153-154 |
| *SAG2.SC2.02.1* | Retrieve Cognitive Map | | 153 |
| *SAG2.SC2.02.2* | Compare Cognitive Maps | | 153 |
| *SAG2.SC2.02.3* | Explore Causes & Effects | | 153-154 |
| *SAG2.SC2.02.4* | Construct Cognitive Map | | 153 |
| *SAG2.SC2.03* | Scenario Building Subsystem | *SAG2.SC3.03, SAG2.SC3.04* | |
| *SAG2.SC2.03.1* | Obtain Trends & Uncertainties from the Knowledgebase | | 154-155 |
| *SAG2.SC2.03.1* | Elicit User's Insights & Opinions | | 154 |
| *SAG2.SC2.03.2* | Generate 3 Scenario Outlines: Best, Worst, Most Likely | | 154 |
| *SAG2.SC2.03.3* | Construct Scenarios using 3 Outlines | | 154 |
| *SAG2.SC2.03.4* | Retrieve Scenario by Name or Keyword | | 154 |
| *SAG2.SC2.03.5* | Modify or Delete Scenario | | 156 |
| | | | |
| *SAG2.SC3* | Domain Knowledge Management System | *SAG2.SC2* | 152-156 |
| *SAG2.SC3.01* | Case Base | | 152-153 |
| *SAG2.SC3.02* | Cognitive Mapping Base | | 153-154 |
| *SAG2.SC3.03* | Scenario Base | | 154-156 |
| *SAG2.SC3.04* | Trend & Uncertainty Base | | 156 |

**Step 9: Your Assessment of the Proposed Solution**

The intent here is that for each Leaf-level Issue, the appropriate *Solution Components* (*SC*s) are identified and an initial assessment of the effectiveness of the given set of *SC*s in addressing the given *Issue*. The Logical Informed Argument method, based on the student's understanding of the actions performed by the given set of *SC*s (e.g. Table 10b), would be used for this initial assessment of Effectiveness. This is why it is important for the student to describe his/her understanding of each *SC* in her/his own words.

**Table 11: Example - Reviewer's Assessment of the Proposed Solution**

| Leaf-level Issue | Solution Component(s) | Your Assessment of Effectiveness |
|---|---|---|
| I01.1: Surface Beliefs | *SAG2.SC2.01:* Case Memory Subsys; *SAG2.SC2.02:* Cognitive Mapping Subsys | Reasonably Effective given features of relevant SCs |
| I01.2: Detect Outmoded Model | *Possibly SAG2.SC2.01 (*Case Memory Subsys) & *SAG2.SC2.02 (*Cognitive Mapping Subsys) | Not mentioned in SAG1; Not clear which features relevant SCs of SAG2 are addressing this Issue |
| I01.3: Unlearn Outmoded Models | *Possibly SAG2.SC2.01 (*Case Memory Subsys) & *SAG2.SC2.02 (*Cognitive Mapping Subsys) | Not mentioned in SAG1; Not clear which features relevant SCs of SAG2 are addressing this Issue |
| I01.4: Enrich mental models | *SAG2.SC2.02:* Cognitive Mapping Subsys; *SAG2.SC2.03:* Scenario Building Subsys | Reasonably Effective given features of relevant SCs |
| I01.5: Validation & Integration | *SAG2.SC2.02:* Cognitive Mapping Subsys; *SAG2.SC2.03:* Scenario Building Subsys | No Cognitive Aid of SAG1; may be partially addressed by some relevant SCs of SAG2 seem to be addressing this Issue |
| I02: Support Creativity | *SAG2.SC2.03:* Scenario Building Subsys | Reasonably Effective given features of relevant SCs |
| I03: Support Forward Thinking | *SAG2.SC2.01:* Case Memory Subsys; | Reasonably Effective given features of relevant SCs |
| I04: Support Backward Thinking | *SAG2.SC2.03:* Scenario Building Subsys | Reasonably Effective given features of relevant SCs |

**Step 10: Analyze the Authors' Evaluation of the Proposed Solution**

Earlier in the paper we presented a list of previously proposed Evaluation Methods for DSR artifacts. Our aim in this step is for the student to increase their understanding of these Evaluation Methods including when each should be applied, and how to appropriately apply each. We also want them to be aware that for their future DRS projects the evaluation of the artifact should cover each Issue that is included in the scope of their *FRP*.

Here the student would first consider and analyze the appropriateness of the evaluation methods that were actually used by the authors (see Table 12a). Material such as that provided in Table 5b as well guidance from the research literature (e.g. Venable, Pries-Heje & Baskerville, 2016) could be used to inform the student's analysis. Next the student would be required to assess the effectiveness of the results of the authors' evaluation (e.g. Figures 7a & 7b) with respect to each leaf-level Issue of the *FRP* (e.g. see Table 12b which covers a subset of the leaf-level *Issues*). In the "Selection" column the student is to discuss whether the selected Evaluation Method is an appropriate choice for evaluating the given artifact; in the "Application" column the student is to discuss whether the authors appropriately applied the selected Evaluation Method.

**Table 12a: Example - Reviewer's Assessment of the Selected Evaluation Methods**

| Solution Artifact Group | Selected Evaluation Method(s) | Your Assessment of Appropriateness of: | |
|---|---|---|---|
| | | Selection | Application |
| SAG2: Architecture | Expert Evaluation | Appropriate: Has been previously used for evaluating architecture artifacts. | Moderate: selection of likely users; some Issues (i.e. Solution Objectives) were not evaluated; selected participants were not sufficiently knowledgeable to assess all aspects. |
| | Logical Informed Argument | Appropriate: Has been previously used for evaluating architecture artifacts. | Moderate: Appropriate to comparison with DSS, but some Issues (i.e. Solution Objectives) were not evaluated |
| SAG3: Instantiation | Expert Evaluation based on a Case | Appropriate for preliminary evaluation of an instantiation artifact | Moderate: selection of likely users; some Issues (i.e. Solution Objectives) were not evaluated; selected participants were nor sufficiently knowledgeable to assess all aspects. |

**Table 12b: Example - Reviewer's Assessment of the Author's Evaluation of the Issues**

| Leaf –level Issue | Adequately Evaluated? | Your Description of the Results |
|---|---|---|
| I01.1: Surface Beliefs | Partly: Expert evaluation; Informed Arguments | *Very Good* to *Excellent* based on Experts' responses to Q7 & Q8 of Table 2; Q.11 Table 3 |
| I01.2: Detect Outmoded Model | No | Not Applicable (N/A) |
| I01.3: Unlearn Outmoded Models | No | N/A |
| I01.4: Enrich mental models | Partly: Expert evaluation | *Very Good* to *Excellent* based on Experts' responses to Q6 - Q8 of Table 2 |
| I01.5: Validation & Integration | No | N/A |
| I02: Support Creativity | Partly: Expert evaluation | *Very Good* based on Experts' responses to Q6 – Q10 of Table 2 |
| I03: Support Forward Thinking | Partly: Expert evaluation | Moderate: p.156: "*All interviewees agreed that the case retrieval and scenario building tools were of great value to most business executives*"; p.157: "… *none of the participants in the case studies had experience with the scenario planning technique. This lack of experience might have reduced these participants' appreciation of the aids provided by the system*" |
| I04: Support Backward Thinking | Partly: Expert evaluation | *Very Good* based on Experts' response to Q4 & Q5 of Table 2 |
| I05: Support Problem Recognition | Partly: Expert evaluation | *Moderate* to *Very Good* based on Experts' response to Q9 & Q11 of Table 2 |
| I13.1: Ease-of-Use | Partly: Expert evaluation | *Very Good* to *Excellent* based on Experts' response to Q1 & Q2 of Table 2 |
| I13.2: Perceived Usefulness | Partly: Expert evaluation | Moderate: p.159: "*the consistency check in the scenario subsystem requires the user to answer a series of questions, which is a cognitively demanding task on the user*") |
| I13.4: Responsiveness | Partly: Expert evaluation | Moderate: (p.159: "*When the number of cases increases dramatically, the searching time for a case is likely to become quite long*") |

> **Your Assessment of the Effectiveness of the Solution based on the Authors' Evaluation:**
> {Describe in your own words + Extract(s) of relevant Verbiage from the paper}
>
> *The solution to offer a promising solution to the FRP as it appears to adequately address most of the Issues (i.e. Solution Objectives) but there are some Issues that do not appear to have been adequately addressed (e.g. I13.2: Perceived Usefulness) while others have not been evaluated.*

**Figure 7a: Results of Authors' Evaluation of the Instantiation Artifact - Expert Opinion Method**

Table 2
Average ratings for CDSS performance

| Questionnaire items | Average ratings | S.D. |
| --- | --- | --- |
| 1. The system is easy to start up. | 4.71 | 0.487 |
| 2. The system is user-friendly. | 4.10 | 0.377 |
| 3. The system's response time is fast. | 4.57 | 0.534 |
| 4. With the Case Memory module, I can quickly record and retrieve business cases, personal experience, etc. | 4.30 | 0.487 |
| 5. Case Memory (reviewing past cases) helps me make better decision. | 3.43 | 0.534 |
| 6. Case Memory helps me find creative solutions to my problems. | 4.00 | 0 |
| 7. The cognitive map may help me think more clearly about the causal relationships among interacting factors concerning a particular policy issue. | 4.43 | 0.534 |
| 8. With the Cognitive Mapping module, I can easily create, retrieve, and modify causal maps. | 4.10 | 0.690 |
| 9. With the Cognitive Mapping module, I explored more interacting factors than I do without it. | 3.57 | 0.534 |
| 10. With the Scenario module, I can construct scenarios quickly. | 3.78 | 0.698 |
| 11. Going through the scenario building process helps me gain a better understanding of general business environments. | 3.85 | 0.690 |
| Average rating | 4.10 | 0.506 |

Source: Chen, J. & Lee, S. (2003)

**Figure 7b: Results of Authors' Evaluation of the Architecture - Informed Argument Method**

Table 3
Comparison between the prototype CDSS and current DSS/ESS

| Major ESS features | Explanations | Typical DSS/ESS | CDSS |
|---|---|---|---|
| 1. User-friendly interface | | X | X |
| 2. Short response time | | X | X |
| 3. Access to aggregate information | | X | |
| 4. Drill down capability | Producing information at various levels of details | X | |
| 5. Critical success factors | Information organized around critical factors vital for attaining the organization's goals | X | |
| 6. Exceptional reports | | X | |
| 7. Trend analysis | Graphs shows trend, ratios, and deviations | X | |
| 8. What-if analysis | Simulation analysis | X | |
| 9. Forecasting | Quantitative forecasting, e.g., sales forecasting | X | |
| 10. Access to Case Memory | Record and retrieve business cases, lesson learned, personal experience, rumors, and speculations | | X |
| 11. Modeling assumptions | Surfacing, representing, comparing assumptions and beliefs. Organizing and synthesizing individual knowledge. | | X |
| 12. Scenario building | Modeling future business environments | | X |

Source: Chen, J. & Lee, S. (2003)

**Step 11: Identify Research Opportunities**

In a given paper, there may be research opportunities to improve the quality of the *FRP* (using insights from Step 6), and/or improve the proposed solution (Steps 9 & 10), and/or improve the evaluation of the proposed solution (Step 10). Table 13 provides examples of research opportunities that could be identified by a student.

**Table 13: Examples of Research Opportunities**

| Step | Research Opportunity | Required Skills, Knowledge, Resources |
|---|---|---|
| 6 | Address *I15.5: Scalability & Flexibility* | **Skills:** Application Development, Database Systems Development<br>**Knowledge:** Cognitive Science, Decision Support Systems<br>**Resources:** Hardware and software resources to build the prototype system |
| 9 | Provide adequate support for addressing Issue *I01.3 (i.e. Unlearn Outmoded Models)*: No Cognitive Aids of SAG1 or features relevant SCs of SAG2 seem to be addressing this Issue | **Skills:** Application Development, Database Systems Development, Interface Design<br>**Knowledge:** Cognitive Science, Organizational Behavior, Decision Support Systems, HCI<br>**Resources:** Hardware and software resources to build the prototype system, Access to real-world executives. |
| 10 | Improve how Issue *I13.2* (*Perceived Usefulness*) is addressed | **Skills:** Application Development (to develop the prototype system), Interface Design<br>**Knowledge:** ICT Performance Issues & Solution Approaches, Managerial Work, DSS & GDSS, HCI |

|  |  | **Resources:** Hardware and software resources to build the prototype system, Access to real-world executives. |

**Step 12: Describe Knowledge Gained from Reviewing this Paper**

For each paper that is reviewed the student document his/her understanding of the knowledge that he/she has gained with respect to Issues, Solutions, and Evaluation Methods aspects. Table 14 provides for each Knowledge Area, an example of what a student might report as having learnt.

**Table 14: Examples - Documentation of Knowledge Gained**

| Knowledge Area | Description |
|---|---|
| *Issues:* | o Mitigate Judgmental Errors (e.g. Availability, Anchoring, Prior Hypothesis biases) |
| *Solutions:* | o Cognitive Map: What it is, and how is can be used |
| *Evaluation Methods* | o Expert Evaluation of Instantiation artifact in the context of a Case Study |

## 5. CONCLUSION

In this paper we presented a structured knowledge acquisition and reflection oriented *Paper Reviewing Process* for evaluating DSR papers. Our paper reviewing process (i.e. Figure 4) and associated conceptual model (i.e. Figure 2) belong to the Design & Action Theory category of Gregor (2006). The approach involved in its development can be considered to be consistent with the DSRM of Peffers at al (2008) as outlined in Table 15 below.

**Table 15: Comparison of Paper Reviewing Process and DSRM Activities**

| DSRM Activity | Presented Paper Reviewing Process |
|---|---|
| Identify Problem & Motivate: Define Problem | The development of reviewing process for DSR papers that involves the student doing a detailed analysis and dissection of the given DSR paper in order for him/her to obtain a deep understanding of the given research problem, its proposed solution artifact(s), and the evaluation of the artifact(s). |
| Define Objectives of a Solution | These are described in Section 1 - *Introduction*. |
| Design & Development | This is described in Section 3 - *Description of the Proposed Paper Reviewing Process.* |
| Demonstration | This is done in Section 4 - *Illustrative Example*. |

| Evaluation | Indirect evaluation evidence include.
|  | o *Unsolicited feedback*:
|  |     ▪ Over the years several students have encouraged me to write a paper on this process as they considered it to have been valuable to their intellectual development, despite its relatively heavy demands compared to other approaches to reviewing papers.
|  |     ▪ Over the years the process has also been revised to address explicitly and implicitly expressed concerns.
|  | o *DSR Journal Papers*: Several DSR papers in reputable journals have been produced by relevant students (e.g. Rao & Osei-Bryson, 2007; Barclay, 2014; Li, Thomas & Osei-Bryson, 2017). |

Hevner and vom Brocke (2023) in a recent paper presented a model that focused on DSR Education including at the doctoral level. Their approach focused on the development of 6 proficiencies. It should be noted that it did not aim to cover all 3 objectives of this paper. In the table we show that our *DSR Paper Reviewing Process* addresses these 6 proficiencies. The reader may note that the *DSR Paper Reviewing Process* presented in this paper was developed before the model of Hevner and vom Brocke (2023).

**Table 16: Connecting the Model of Hevner & vom Brocke & The DSR Paper Reviewing Process**

| Model of Hevner & vom Brocke (2023) | DSR Paper Reviewing Process |
|---|---|
| Representing the Problem Space (*Proficiency P1*):<br><br>"*The first DSR proficiency requires that students be able to define and represent a tractable Problem Space for the DSR project*" | Steps 0 – 5 including:<br><br>o Step 4: *Define the Formal Research Problem (FRP)*<br>o Step 5: *Define the Informal Research Problem (IRP)* |
| Capturing Extant Knowledge in the Solution Space (*Proficiency P2*):<br><br>"*The key challenge in this proficiency is to capture all applicable knowledge from both the technical and scientific knowledgebases to effectively perform the DSR project*" | o Step 5: *Describe the IRP* - Review and analyze your narrative description of the IRP in order to determine any relevant previously known IS *Issue* (***K***) that are included in the scope of the *Informal Research Problem* (***IRP***) |
| Controlling the DSR Process (*Proficiency P3*):<br><br>"Several proposed process models for scheduling and coordinating design activities exist" | Figure 4 (Description of the Paper Reviewing Process) describes a sequence of integrated activities |

| Building Innovative Design Artifacts (*Proficiency P4*): | Steps 8 & 9 should increase the user's understanding with respect to designing appropriate solutions. |
| --- | --- |
| | o Step 8: *Provide your Description of the Proposed Solution* |
| | o Step 9: *Provide your Assessment of the Authors' Proposed Solution* |
| Measuring the Satisfaction of Research Goals With Rigorous Evaluation (*Proficiency P5*) | o Step 10: *Analyze the Authors' Evaluation of the Proposed Solution* |
| Contributing to Science and Practice (*Proficiency P6*) | o Step 11: *Identify Research Opportunities* |
| | o Step 12: *Describe Knowledge Gained* |

**5.1 Contributions to ICT4D Academics**

Our knowledge acquisition and reflection oriented *Paper Reviewing Process*, has as its major objectives: 1) Increasing the ICT4D researcher's content knowledge of the given subject; 2) Increasing the ICT4D researcher's knowledge and understanding of DSR processes, and 3) Facilitating the Identification of opportunities for new DSR projects.

1. *Increasing the ICT4D researcher's content knowledge of the given subject*: In Step 12 the user would document the new knowledge gained from the detailed dissection-oriented review of the given paper. By documenting the new knowledge obtained from the review of each of the selected papers, the user would have a record of what he/she has learned
2. *Increasing the ICT4D researcher's knowledge and understanding of the DSR process*: This is primarily accomplished by the user engaging in Steps 8 - 10 of the process.
3. *Facilitating the Identification of opportunities for new DSR projects*: In Step 11 the user would assess and document opportunities for new DSR projects.

Our paper reviewing process also offers the opportunity to increase the number of ICT4D researchers who can serve competently and comfortably as reviewers of DSR papers, even if they choose to use a modified version of the presented paper reviewing process**.**

**5.2 Contributions to ICT4D Practice**

This research has the potential to offer several contributions to practitioners. Two of these will be discussed below.

1. As noted earlier, most ICT artifacts (e.g. including hardware, software, models, methodologies, frameworks) that are used in countries with developing economies were designed and developed in countries with advanced economies and typically are based on the values and perspectives of such societies.  There is thus the need to ensure that any acquired artifact has an adequate contextual fit or can be effectively modified to have such a fit. An implication of this is that the ICT4D practitioner would to need to develop an intimate level of understanding of the given ICT4D application problem, and also to quickly develop an appropriate level of understanding with respect to the considered artifacts, including their the associated design objectives, assumptions, limitations, etc.  Even a modified application of Steps 4 & 5 could be beneficial for the ICT4D practitioner developing such an intimate level of understanding of the given ICT4D application problem, and also of the artifact's designers design objectives, assumptions, limitations.  Similarly, even a modified application of Steps 9 & 10, would be beneficial to the ICT4D practitioner developing a reasonable assessment as to whether each considered set of artifacts offers an adequate fit for the requirements of the given ICT4D application problem.
2. The achievement of objectives (1)-(3) of the contributions to academics would also be beneficial to practitioners as some of these practitioners may currently be or have been students taught by these academics.  Such academics increasing knowledge, competence, and confidence with respect to technical areas of IS and also DSR could play out in the classroom in terms of more technical IS courses that are appropriate for the given ICT4D context.

**5.3 Limitations & Future Research**

We have presented a new *Paper Reviewing Process* that is partly based on a single DSR process model, that of Peffers at al (2008). The reason for this is that this DSRM is similar to the systems analysis & design methodologies that researchers and practitioners would like have encountered even in an undergraduate IS program. It is possible that another DSR process model could have been used, but what is also important is that *Paper Reviewing Process* was developed that offers benefits to ICT4D academics and practitioners.

This paper is presented to the community with the hope that it will attract other researchers and educators to use, critique and improve on the presented paper reviewing process, thus increasing the likelihood for its successful contribution to the development of DSR-oriented ICT4D researchers.

## REFERENCES

1. Andoh-Baidoo, F., Osatuyi, B., & Kunene, K. N. (2014). Architecture for managing knowledge on cybersecurity in Sub-Saharan Africa. *Information Technology for Development*, *20*(2), 140-164.
2. Bailey, A., & Osei-Bryson, K. M. (2018). Contextual reflections on innovations in an interconnected world: Theoretical lenses and practical considerations in ICT4D. *Information Technology for Development*, *24*(3), 423-428
3. Barclay, C. (2014). Using frugal innovations to support cybercrime legislations in small developing states: introducing the cyber-legislation development and implementation process model (CyberLeg-DPM). *Information Technology for Development*, *20*(2), 165-195.
4. Asante Boakye, E., Zhao, H., & Ahia, B. N. K. (2022). Blockchain technology prospects in transforming Ghana's economy: a phenomenon-based approach. *Information Technology for Development*, 1-30.
5. Bunnell, L., Osei-Bryson, K.-M., & Yoon, V. (2019). RecSys issues ontology: a knowledge classification of issues for recommender systems researchers. ***Information Systems Frontiers***, 1-42, in press. https://doi.org/10.1007/s10796-019-09935-9
6. Chen, J. & Lee, S. (2003). An exploratory cognitive DSS for strategic decision making. ***Decision Support Systems*, *36*(2)**, 147-160.
7. Delone, W. H., & McLean, E. R. (2003). The DeLone and McLean Model of Information Systems Success: A Ten-Year Update. ***Journal of Management Information Systems*, *19*(4)**, 9-30.
8. Drechsler, A., & Hevner, A. (2016). A Four-Cycle Model of IS Design Science Research: Capturing the Dynamic Nature of IS Artifact Design. In *Breakthroughs and Emerging Insights from Ongoing Design Science Projects: Research-in-progress papers and poster presentations from the **11th International Conference on Design Science Research in Information Systems and Technology (DESRIST) 2016***. *St. John, Canada, 23-25 May*. DESRIST 2016.
9. Gregor S. and Jones D. (2007) The Anatomy of a Design Theory. ***Journal of The Association For Information Systems 8(5)***, 312-335.
10. Hoefsloot, F. I., Jimenez, A., Martinez, J., Miranda Sara, L., & Pfeffer, K. (2022). Eliciting design principles using a data justice framework for participatory urban water governance observatories. *Information Technology for Development*, *28*(3), 617-638.
11. Hevner, A., March, S., Park, J., & Ram, S. (2004). Design Science in Information Systems Research. ***Management Information Systems Quarterly, 28(1)***, 75-105.
12. Hevner, A. & vom Brocke, J. (2023) A Proficiency Model for Design Science Research Education. Journal of Information Systems Education, 34(3), 264-278.
13. Hirschheim, R. (2008). Some Guidelines for the Critical Reviewing of Conceptual Papers. ***Journal of the Association for Information Systems*, *9*(8)**, 21.

14. Hosack, B., Hall, D., Paradice, D., & Courtney, J. F. (2012). A Look Toward The Future: Decision Support Systems Research is Alive and Well. Journal of the Association for Information Systems, 13(5), 315.
15. Kwan, M. M., & Balasubramanian, P. (2003). KnowledgeScope: Managing Knowledge in Context. **Decision Support Systems, 35(4)**, 467-486.
16. Lee, A. S. (1995). Reviewing a Manuscript for Publication. **Journal of Operations Management, 13(1)**, 87-92.
17. Li, Y., Thomas, M. A., & Osei-Bryson, K. M. (2017). Ontology-based data mining model management for self-service knowledge discovery. **Information Systems Frontiers, 19(4)**, 925-943.
18. Li, Y., Thomas, M. A., Stoner, D., & Rana, S. S. (2020). Citizen-centric capacity development for ICT4D: the case of continuing medical education on a stick. **Information Technology for Development, 26(3)**, 458-476.
19. Maes, A., & Poels, G. (2006, November). Evaluating quality of conceptual models based on user perceptions. In **International Conference on Conceptual Modeling** (pp. 54-67). Springer, Berlin, Heidelberg.
20. March, S. T., & Smith, G. F. (1995). "Design and Natural Science Research on Information Technology", **Decision Support Systems, 15(4)**, 251-266.
21. Mullarkey, M. T., & Hevner, A. R. (2019). An Elaborated Action Design Research Process Model. *European* **Journal of Information Systems, 28(1)**, 6-20.
22. Mushi, G. E., Serugendo, G. D. M., & Burgi, P. Y. (2023). Data management system for sustainable agriculture among smallholder farmers in Tanzania: research-in-progress. **Information Technology for Development,** 1-24.
23. Niederman, F., & March, S. T. (2012). Design Science and the Accumulation of Knowledge in the Information Systems Discipline. **ACM Transactions on Management Information Systems (TMIS), 3(1)**, 1.
24. Namyenya, A., Daum, T., Rwamigisa, P. B., & Birner, R. (2022). E-diary: a digital tool for strengthening accountability in agricultural extension. **Information Technology for Development, 28(2)**, 319-345.
25. Osei-Bryson, K. M., & Bailey, A. (2019). Contextual reflections on innovations in an interconnected world: theoretical lenses and practical considerations in ICT4D–Part 2. **Information Technology for Development, 25(1)**, 1-6.
26. Osei-Bryson, K. M., Brown, I., & Meso, P. (2022). Advancing the Development of Contextually Relevant ICT4D Theories-From Explanation to Design. **European journal of information systems, 31(1)**, 1-6.
27. Peffers, K., Tuunanen, T., Rothenberger, M. A., & Chatterjee, S. (2007). A Design Science Research Methodology for Information Systems Research. **Journal of Management Information Systems, 24(3)**, 45-77.
28. Peffers, K., Rothenberger, M., Tuunanen, T., & Vaezi, R. (2012). Design Science Research Evaluation. In **International Conference on Design Science Research in Information Systems** (pp. 398-410). Springer Berlin Heidelberg.
29. Qureshi, S. (2020). Why data matters for development? Exploring data justice, micro-entrepreneurship, mobile money and financial inclusion. **Information Technology for Development, 26(2)**, 201-213.
30. Rao, L., & Osei-Bryson, K. M. (2007). Towards Defining Dimensions of Knowledge Systems Quality. *Expert Systems with Applications*, *33*(2), 368-378
31. Rao, L., & McNaughton, M. (2019). A knowledge broker for collaboration and sharing for SIDS: the case of comprehensive disaster management in the Caribbean. **Information Technology for Development, 25(1)**, 26-48.

32. Sein, M. K., Henfridsson, O., Purao, S., Rossi, M., & Lindgren, R. (2011). Action Design Research. *MIS Quarterly 35(1)*, 37-56.
33. Singh, S. K., Jenamani, M., Dasgupta, D., & Das, S. (2021). A conceptual model for Indian public distribution system using consortium blockchain with on-chain and off-chain trusted data. *Information Technology for Development*, *27*(3), 499-523.
34. Stewart, G., & Osei-Bryson, K. M. (2013). Exploration of Factors that Impact Voluntary Contribution to Electronic Knowledge Repositories in Organizational Settings. *Knowledge Management Research & Practice*, **11(3)**, 288-312.
35. Vaishnavi, V. & Kuechler, W. (2004) Design Research in Information Systems. http://desrist.org/design-research-in-information-systems
36. van Biljon, J., Marais, M., & Platz, M. (2017). Digital platforms for research collaboration: using design science in developing a South African open knowledge repository. *Information Technology for Development*, *23*(3), 463-485.
37. Venable, J., Pries-Heje, J., & Baskerville, R. (2016). FEDS: a framework for evaluation in design science research. *European journal of information systems*, **25(1)**, 77-89.
38. Yeoh, W., & Koronios, A. (2010). Critical Success Factors for Business Intelligence Systems. *Journal of Computer Information Systems, 50(3)*, 23-32.

## APPENDIX: Illustrative Example – Issues of the IRP & FRP

The table below presents for our illustrative example the set of Issues in the *FRP* along with a set of Issues that could be considered to be relevant to the scope of the *IRP*. In the case of *K* Issues (which are taken from the community Knowledge-base) in some cases all the child Issues of a given *K* Issue may not have been included.

**Table A1: Issues of the FRP & IRP**

| Label | Issue_Desc | Issue_Text | D/I/U/K | Page # |
|---|---|---|---|---|
| I01 | Support Mental Modeling | cognitive aspect of decision support | D | 147 |
| I01.1 | Surface Beliefs | providing tools to surface DM's tacit assumptions and beliefs | D | 148 |
| I01.2 | Detect Outmoded Model | provide direct aid in detecting and unlearning outmoded mental models | I | 148 |
| I01.3 | Unlearn Outmoded Models | provide direct aid in detecting and unlearning outmoded mental models | I | 148 |
| I01.4 | Enrich mental models | consciously helping enrich the … mental models | D | 149 |
| I01.5 | Validation & Integration | facilitating mental model validation and integration | D | 149 |
| I02 | Support Creativity | …. facilitating … creative thinking | D | 148 |
| I03 | Support Forward Thinking | aid forward thinking of DM | D | 148 |
| I03 | Support Forward Thinking | supporting the decision maker's forward … thinking | D | 149 |
| I04 | Support Backward Thinking | *supporting the decision maker's backward ... thinking* | D | 149 |
| I05 | Support Problem Recognition | *provide support in problem and opportunity recognition and diagnosis* | I | 149 |
| I06 | Mitigate Judgmental Errors | *mitigating judgmental errors due to human limited information processing capabilities* | D | 149 |
| I06.1 | Minimize Availability Bias | *Their limited ability to retrieve past cases may cause them to make biased judgments and decisions* | D | 150 |
| I06.2 | Minimize Anchoring Bias | *When executives plan for the uncertain future, they anchor on past experience* | D | 150 |
| I06.3 | Minimize Prior Hypothesis Bias | *Individuals tend to seek and use information consistent with their beliefs rather than information that is inconsistent.* | D | 150 |
| I06.4 | Minimize Analogy Error | *Reasoning by analogy has also been shown to be effective ... associations between existing circumstances and past events can be inappropriate ...* | D | 150 |
| I06.5 | Reduce Overconfidence | *Overconfidence can be dangerous. It indicates that people often do not know how little they know and how much additional information they need* | D | 151 |
| I07 | Introspective Inquiry | *support introspective mode* | I | 150 |
| I08 | Dialectic Inquiry | *support dialectic inquiry mode* | I | 150 |
| I09 | Eclectic Inquiry | *support eclectic inquiry mode* | I | 150 |
| I10 | Support Group Activities | *a multi-participant decision support system running on the Intranet in an organization* | D | 152 |
| I11 | Security | *handle both private and public messages; can store private domain knowledge* | I | 152 |
| I11.1 | Confidentiality | o *handle ... private ... messages;* <br> o *can store private domain knowledge* | I | 152 |
| I11.2 | Availability | *capable of knowledge selection* | I | 152 |
| I11.3 | Integrity | *refers to the prevention of unauthorized or improper data modification.* | U | KB-Sc |
| I12 | Knowledge Management | *capable of knowledge acquisition, selection, and derivation* | I | 152 |
| I12.1 | Acquisition | *capable of knowledge acquisition* | I | 152 |
| I12.2 | Support Querying | *capable of knowledge selection* | I | 152 |
| I12.3 | Inferencing | *capable of knowledge derivation* | I | 152 |
| I12.4 | Knowledge-base Durability | *can store public domain knowledge* | I | 152 |
| I13 | User Acceptance | *... provide ... users ... easy data access ... a useful conceptual tool* | I | 147-148 |
| I13.1 | Ease-of-Use | o *... provide their users ... easy data access* <br> o *provide ... information ... in a... user-friendly fashion* | I | 147 |

| I13.2 | Usefulness | *… provides a useful conceptual tool for decision makers* | I | 148 |
| I13.3 | User Satisfaction | *"… the extent to which users believe the information system available to them meets their information requirements …'"* | U | KB-UA |
| I13.4 | Responsiveness | *"EIS provide … information … in a timely … fashion "* | I | 147 |
| I14 | Evaluation & Choice | *"… instead of providing support for the evaluation and choice phase of the decision-making process"* | X | 149 |
| I15 | KM System Quality |  | K | KB-KM |
| I15.1 | Currency | *"describes when the information was entered in the system"* | K | KB-KM |
| I15.2 | Credibility | *"describes the credibility of the source that provided the information"* | K | KB-KM |
| I15.3 | Transactional Availability | *"describes the percentage of time the information … is available due to the absence of update processes"* | K | KB-KM |
| I15.4 | Volatility | *"describes the time period for which the information is valid in the real world"* | K | KB-KM |
| I15.5 | Scalabilty & Flexibiliy | *" … easy expansion of the system to align it with evolving information needs"* | K | KB-KM |

KB-KM: Source is the community Knowledge-base (KB) on Knowledge Management
KB-UA: Source is the community Knowledge-base (KB) on User Acceptance
KB-Sc: Source is the community Knowledge-base (KB) on IS Security